\documentclass[conference]{IEEEtran}
\IEEEoverridecommandlockouts
\usepackage{cite}
\usepackage{amsmath,amssymb,amsfonts}
\usepackage{algorithmic}
\usepackage{graphicx}
\usepackage{textcomp}
\usepackage{listings}
\usepackage{xcolor}
\usepackage{todonotes}
\usepackage[normalem]{ulem}
\usepackage{hyperref}

\definecolor{codegreen}{rgb}{0,0.6,0}
\definecolor{codegray}{rgb}{0.5,0.5,0.5}
\definecolor{codepurple}{rgb}{0.58,0,0.82}
\definecolor{backcolour}{rgb}{0.95,0.95,0.92}

\lstdefinestyle{mystyle}{
    backgroundcolor=\color{backcolour},   
    commentstyle=\color{codegreen},
    keywordstyle=\color{magenta},
    numberstyle=\tiny\color{codegray},
    stringstyle=\color{codepurple},
    basicstyle=\ttfamily\footnotesize,
    breakatwhitespace=false,         
    breaklines=true,                 
    captionpos=b,                    
    keepspaces=true,                 
    numbers=left,                    
    numbersep=4pt,                  
    showspaces=false,                
    showstringspaces=false,
    showtabs=false,                  
    tabsize=4,
    basicstyle=\small
}

\newcommand{\pd}[2] {\frac{\partial #1}{\partial #2}}

\def\BibTeX{{\rm B\kern-.05em{\sc i\kern-.025em b}\kern-.08em
    T\kern-.1667em\lower.7ex\hbox{E}\kern-.125emX}}
\begin{document}

\title{GPU-Accelerated Discontinuous Galerkin Methods:\\ 30x Speedup on 345 Billion Unknowns\\
\thanks{This work was partially supported under ONR Grant N00014-16-1-2737.}
}

\author{
  \IEEEauthorblockN{Andrew C. Kirby}
  \IEEEauthorblockA{\textit{Lincoln Laboratory Supercomputing Center} \\
  \textit{Massachusetts Institute of Technology}\\
  Lexington, MA, USA}
\and
  \IEEEauthorblockN{Dimitri J. Mavriplis}
  \IEEEauthorblockA{\textit{Mechanical Engineering} \\
  \textit{University of Wyoming}\\
  Laramie, WY, USA}
}

\maketitle
\begin{abstract}
A discontinuous Galerkin method for the discretization of the compressible Euler equations, the governing equations of inviscid fluid dynamics, on Cartesian meshes is developed for use of Graphical Processing Units via OCCA, a unified approach to performance portability on multi-threaded hardware architectures. A 30$\mathbf{x}$ time-to-solution speedup over CPU-only implementations using non-CUDA-Aware MPI communications is demonstrated up to 1,536 NVIDIA V100 GPUs and parallel strong scalability is shown up to 6,144 NVIDIA V100 GPUs for a problem containing 345 billion unknowns. A comparison of CUDA-Aware MPI communication to non-GPUDirect communication is performed demonstrating an additional 24\% speedup on eight nodes composed of 32 NVIDIA V100 GPUs.
\end{abstract}


\section{Introduction}
A transformation in computing architectures has emerged over the past decade in the quest to keep Moore's Law alive, leading to significant advancements in Artificial Intelligence and Machine Learning techniques\cite{Hoefler:2019,Reuther:2019}. The success of Machine Learning has caused a role reversal, now becoming a primary driver in the development of massively-parallel thread-based hardware indicative of future exascale-era architectures \cite{Messina:2017}. Traditional Computational Science applications such as Computational Fluid Dynamics (CFD) has eschewed the use of Graphical Processing Units (GPU) due to their programming complexity to achieve high performance. However, in this volatile heterogeneous computing landscape, algorithms suitable for these platforms are now forefront in the advancement of these fields.

Low-order discretizations such as the finite volume method \cite{Jameson:1986,Mavriplis:1988} have been the industrial standard in CFD for the last 40 years. However, these algorithms suffer from low arithmetic intensity\footnote{Arithmetic intensity is the ratio of floating-point operations performed to the amount of memory used in the calculation.} and were never concieved with massive parallelism in mind, making them potentially ill-suited for heterogeneous systems containing GPUs. According to the NASA CFD Vision 2030 Study\cite{Slotnick:2014}, "Beyond potential advantages in improved accuracy per  degree of freedom, higher-order methods may more effectively utilize new HPC hardware through increased levels of computation per degree of freedom." To leverage these new architectures and enable higher-fidelity simulations, algorithmic advancement of high-order methods, such as the discontinuous Galerkin method \cite{Hesthaven:2007} and the flux-reconstruction method \cite{Witherden:2014}, have emerged within the Computational Science community. These high-order methods have higher-algorithmic intensity and naturally-tiered parallelism through inter-element distributed memory strategies and intra-element shared memory parallelism.

This work develops a discontinuous Galerkin (DG) method to discretize the governing equations of fluid dynamics concerning aerospace and atmospheric applications. The discontinuous Galerkin method exhibits highly-parallelizable computational kernels due to its nearest-neighbor-only element communication pattern and high-intensity localized compute stencil, making them well suitable for GPUs in a distributed computing environment. Implementation of a DG methods for hyperbolic systems of equations on GPUs was introduced in \cite{Klockner:2009} and extension and optimization of this work conducted in \cite{Gandham:2015, Modave:2016, Chan:2016, Karakus:2016a, Karakus:2016b, Karakus:2019}.

We employ an thread-programming abstraction model known as OCCA \cite{Medina:2014}, an open-source library, enabling the use of GPUs, CPUs, and FPGAs, to achieve performance portability. OCCA allows developers to write performance-portable code in the OCCA Kernel Language (OKL)\cite{Medina:2015} which is an intermediate representation for shared-memory parallelism. The intermidiate representation is just-in-time compiled to various backends including serial code, OpenMP, CUDA, and OpenCL. Efficient and scalable implementations of high-order discretizations for atmospheric conditions using OCCA have been demonstrated by Abdi et. al in \cite{Abdi:2019a} and \cite{Abdi:2019b}.

A nodal DG method with collocation of integration and solution points is assumed enabling several arithmetic simplifications. Further, we conform to Cartesian meshes resulting in constant mesh parameters further simplifying the discretization. This is done to establish a performance ceiling for future algorithmic developments involving dynamically adapting unstructured meshes and variable polynomial order element-wise solutions, i.e. $hp-$adaption. All computations involved in the discretization of the governing equations are ported to the GPU for this work including spatial and temporal residuals. The performance of the method is analyzed and comparisons of network communications via CUDA-Aware MPI are performed. Lastly, we demonstrate the parallel scalability of the implementation up to 6,144 NVIDIA V100 GPUs.

\section{Discontinuous Galerkin Method}
\subsection{Governing Equations}
The governing equations utilized in this work are the 
three-dimensional compressible inviscid Euler equations, which are written in 
conservative form:

\begin{equation}
\pd{\textbf{Q}\left(\boldsymbol{x},t\right)}{t} + 
\vec{\nabla} \cdot \textbf{F} \left(\textbf{Q}\left(\boldsymbol{x},t\right)\right) = 0
\end{equation}
representing the conservation of mass, momentum, and energy for a 
fluid. The solution vector $\textbf{Q}$ represents the conservative flow 
variables and the matrix $\textbf{F}$ represents the Cartesian flux 
components. $\textbf{Q}$ and $\textbf{F}$ are defined as follows:

\begin{equation} \label{NS-equations}
\begin{split}
\begin{array}{l}
 \ \ \ \ \ \ \ \ \ \ \ \ \ \ \ \ \ \ \ \ \ \ \ \ \ \ \ \ \ 
 \underline{\textbf{F}^1}
 \ \ \ \ \ \ \ \ \ 
 \underline{\textbf{F}^2}
 \ \ \ \ \ \ \ \ \ 
 \underline{\textbf{F}^3} \\
 \textbf{Q} = 
   \left\{ 
     \begin{array}{c} 
       \rho \\ \rho u \\ \rho v \\ \rho w \\ \rho E  \\ 
     \end{array}
   \right\},  
 \textbf{F} =
 \left\{
 \begin{array}{lll}
 	\rho u       &\rho v       &\rho w       \\
 	\rho u^2 + p &\rho u v     &\rho u w     \\
 	\rho u v     &\rho v^2 + p &\rho v w     \\
 	\rho u w     &\rho v w     &\rho w^2 + p \\ 
 	\rho u H     &\rho v H     &\rho w H 
 \end{array} 
 \right\}
 \end{array}
 \end{split} 
\end{equation}
where $\rho$ is density, $u,v,w$ are velocity components in 
each spatial coordinate direction, $p$ is pressure, $E$ is total internal 
energy, and $H = E + \frac{p}{\rho}$ is total enthalpy. 
These equations are closed using the ideal gas equation of state:
\begin{equation}
\rho E = \frac{p}{\gamma -1} + \frac{1}{2}\rho \left( u^2 + v^2 + w^2 \right)
\end{equation}
where $\gamma = 1.4$ is the ratio of specific heats.

\begin{figure*}[ht]
\centerline{\includegraphics[width=1.0\linewidth]{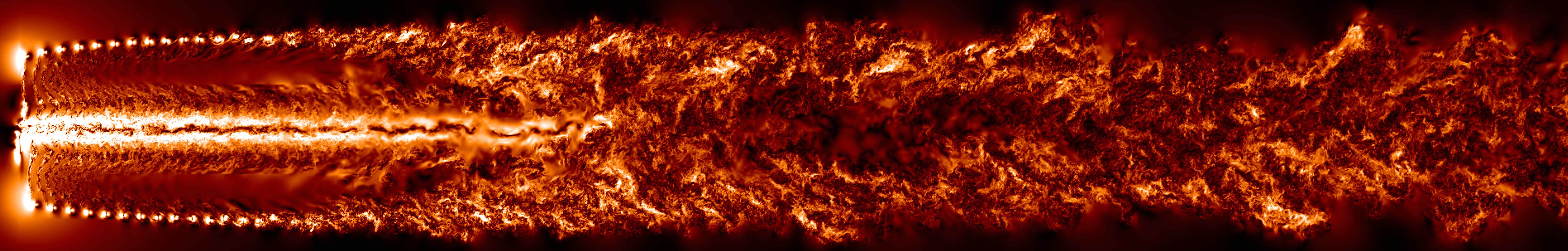}}
\caption{Turbulent wake structure of a wind turbine containing 1.6 billion degrees-of-freedom (8 billion unknowns). This simulation used the CPU implementation of this work on 20,000+ CPU cores.}
\label{fig:wake}
\end{figure*}

\subsection{Discretization}
This work constrains the computational domain to Cartesian meshes. 
Under these assumptions, arithmetic simplifications are performed to increase computational performance.
Within this setting, the governing equations are discretized via the discontinuous Galerkin (DG) method \cite{Cockburn:2012} assuming a weighted residual formulation. A weak formulation is established by taking the inner product of the governing equations and set of piecewise continuous test functions, where the inner product is defined as multiplying each equation by a set of test functions $\psi_s \in l(x)$, the Lagrange basis polynomials up to degree $N$, and integrating over the Cartesian element volume $\Omega_k$:

\begin{equation}
\int_{\Omega_k}  
\left( 
\pd{\textbf{Q}}{t} + \vec{\nabla} \cdot \textbf{F} 
\right) 
\psi_{s}(\boldsymbol{x}) \mathrm{d}\boldsymbol{x} = 0 \label{DG-Integral}
\end{equation}
Integrating \eqref{DG-Integral} by parts, the weak-form 
residual $\textbf{R}_{s}^{\text{Weak}}$ is defined as:

\begin{equation}    \label{weakform-resid}
\begin{split}
\textbf{R}_{s}^{\text{Weak}} &= 
 \int_{\Omega_k}  
  \pd{\textbf{Q}}{t} 
      \psi_{s}(\boldsymbol{x})\ \mathrm{d}\boldsymbol{x} \\
 &- \int_{\Omega_k} 
   \left( \textbf{F} \cdot \vec{\nabla} \right) 
		 \psi_{s}(\boldsymbol{x})\ \mathrm{d}\boldsymbol{x} \\
 &+ \int_{\Gamma_k} 
   \left( \textbf{F}^{*} \cdot \vec{\mathrm{n}} \right) 
	 	 \psi_{s}(\boldsymbol{x}|_{\Gamma_k})\ \mathrm{d} \Gamma_k = 0
\end{split}
\end{equation}
The discrete residual contains integrals over mesh element boundaries $\Gamma$ where special 
treatment is needed for the fluxes. The advective fluxes 
$\textbf{F}^*$ are calculated using an approximate Riemann solver, namely the Lax-Friedrichs method \cite{Harten:1983}. 
To complete the spatial discretization, the solution is approximated as follows:
\begin{equation}
\textbf{Q}(x,t) = \sum\limits_{s=1}^N \hat{Q}_s(t) \phi_s(x)
\end{equation}
where $\phi_s(x)$ are chosen to be the same as the test functions $\psi_s(x)$.
A collocation approach is chosen for computational efficiency by choosing the basis and test functions to be Lagrange interpolation polynomials  
\begin{equation}\label{Nodal}
\ell_s \left(x\right) = 
\prod_{i=1,i \neq s}^N 
\frac{\left( x - \xi_i \right)}{\left( \xi_s - \xi_i \right)}, 
\ \ \ s=1,\ldots,N
\end{equation}
where $N=p+1$: $p$ is the user-chosen polynomial degree, and numerical integration via Gauss-Legendre quadrature is used. 
Lastly, we implement a low-storage explicit Runge-Kutta method to discretize the temporal component of \eqref{DG-Integral}.

\section{CPU Implementation}
This work serves within the larger computational framework known as the \textbf{W}yoming \textbf{W}ind and \textbf{A}erospace \textbf{A}pplications \textbf{\textit{K}}ompution \textbf{E}nvironment (WAKE3D) \cite{Brazell:2016,Kirby:2017,Kirby:2018a}. WAKE3D has been demonstrated on various aerodynamics problems \cite{Kara:2020} and wind energy applications \cite{Kirby:2019,Edmonds:2019,Kirby:2018,Kirby:2016}. The framework is composed of multiple software components, namely NSU3D \cite{Mavriplis:2015}, DG4est \cite{Kirby:2019,Stoellinger:2019}, and TIOGA \cite{Brazell:2016:tioga,Sitaraman:2020}. Figure~\ref{fig:wake} illustrates the simulation of a Siemens SWT-2.3-93 wind turbine using WAKE3D.

\begin{figure}[ht]
\centerline{\includegraphics[scale=0.225]{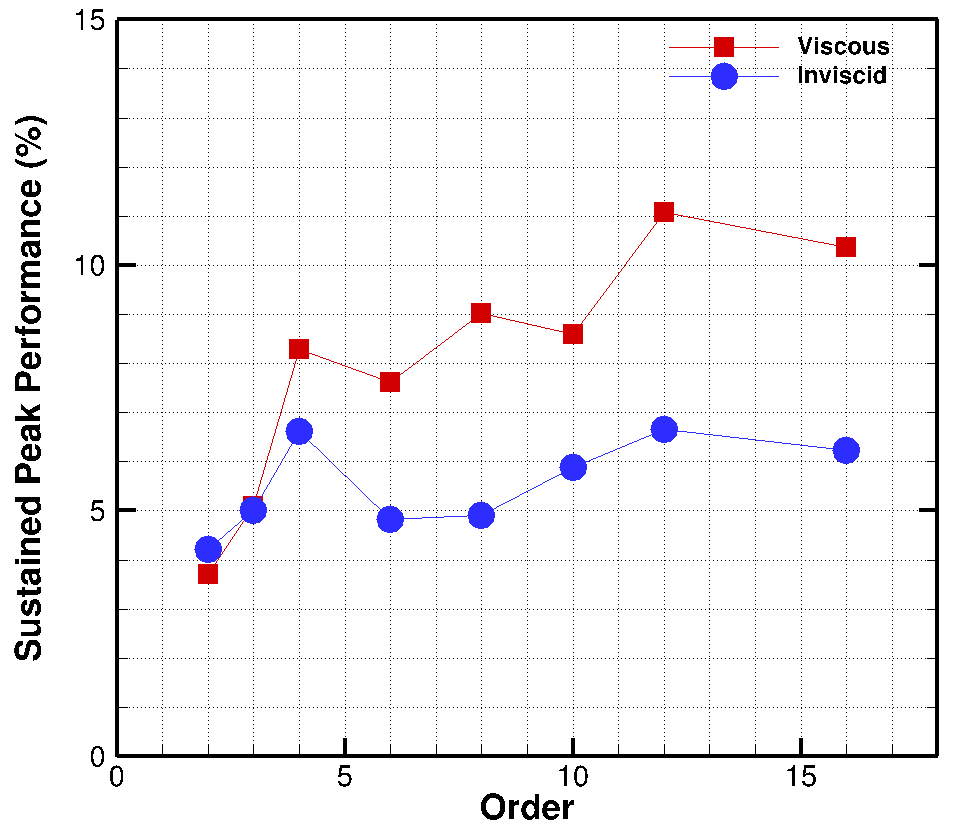}}
\caption{Sustained peak performance of CartDG on Intel Xeon E5-2697V2 processors for various solution orders.}
\label{fig:peak}
\end{figure}

The computational kernel within the DG4est component, known as CartDG \cite{Kirby:2015,Brazell:2015,Kirby:thesis}, is the primary focus of this work. CartDG is a discontinuous Galerkin method designed for computational efficiency on Cartesian meshes utilizing tensor-product collocation-based basis functions. The CPU implementaion has achieved over 10\% sustained peak of theoretical compute performance \cite{Brazell:2017} using the viscous formulation as shown in Figure~\ref{fig:peak}. CartDG utilizes the Message Passage Interface (MPI) for distributed-memory computation and enables computation-communication overlap to hide communication latency. It has been demonstrated to scale to over one million MPI ranks \cite{Brazell:2016} on ALCF Mira Supercomputer.

\section{GPU Implementation}
All computational kernels herein are ported via OCCA including the spatial volume and surface residuals and a five-stage low-storage Runge-Kutta temporal discretization. Discontinuous Galerkin methods exhibit multiple levels of parallelism: coarse-grain parallelism via mesh decomposition and fine-grain parallelism via single instruction, multiple data within a mesh element where element-local degrees-of-freedom are coupled. In this work, the MPI programming model is employed for coarse-grain parallelism and OCCA \cite{Medina:2014} is utilized for fine-grain parallelism. We select the CUDA-backend within OCCA to execute all computational kernels on NVIDIA V100 GPUs. OCCA provides a kernel language known as OKL, a simple extension of the C-language, enabling explicit architecture-independent fine-grain parallelized code. The sample code shown illustrates the spatial residual volume kernel (second term) from \eqref{DG-Integral} using OKL. Note the similarity of the OKL-based code to the C programming language interlaced with simple decorators, e.g. @kernel, @exclusive, @shared, @outer, @inner.

The discontinuous Galerkin method discretizes mesh elements into multiple solution modes per variable which depend on the solution polynomial degree established a priori by the user. For example, in three-dimensional space, suppose the polynomial degree $p=6$, then the number of modes per conservative variable within an element is $(p+1)^3 = 343$. To map the discretization to the hardware, we assign one hardware thread per mode, which is seen in the code sample indicated by the \verb+cubeThreads+ definition. The \verb+@inner+ attribute assigns a thread to each index $(i,j,k)$ mode.

\begin{lstlisting}[float=*, language=C, caption=OCCA Volume Flux Residual Kernel Example]
#define cubeThreads                            \ 
  for (int k = 0; k < p_TM; ++k; @inner(2))    \ 
    for (int j = 0; j < p_TM; ++j; @inner(1))  \ 
      for (int i = 0; i < p_TM; ++i; @inner(0))

/* Inviscid Flux Volume Kernel--Total Operations: nelements*(38*N^3 + 30*N^4) */
@kernel void residual_inviscid_volume_occa(
						  	/* solution */ @restrict const double *Q,
					    	/* residual */ @restrict double *R,
				      /* basis functions*/ @restrict const double *dphiw, 
			      /* mesh element count */ const int nelements,
   /* ratio of specific heats minus one */ const double gm1,
                     /* geometry factor */ const double vgeo){
    for (int e = 0; e < nelements; ++e; @outer) {
        const int tm_nf     = p_TM*p_NF; 	 // p_NF: # of fields = 5
        const int tm_tm_nf  = p_TM*tm_nf;	 // p_TM: # of modes = p+1 
        const int nf_tm_ALL = p_TM*tm_tm_nf;
        const int elem_idx  = e*nf_tm_ALL;

        @exclusive double r_Q[p_NF]; // thread-local solution
        @exclusive double r_R[p_NF]; // thread-local residual

        @shared double s_fluxX[p_NQP][p_NQP][p_NQP][p_NF]; //s_fluxY, s_fluxZ
        @shared double s_dphiw[p_TM][p_TM]; // basis with quadrature weights

        // load solution and reset residual per thread memory
        cubeThreads {
            const int idx1 = i*p_NF + j*tm_nf + k*tm_tm_nf + elem_idx;
            #pragma unroll p_NF
            for (int n = 0; n < p_NF; ++n) {
                r_Q[n] = Q[idx1+n];
                r_R[n] = 0.0;
            }
            // fetch transpose(dphiw) to shared memory
            if(k == 0) s_dphiw[i][j] = dphiw[j*p_TM+i];
        } @barrier("localMemFence");

        // volume flux: ops = 38*N^3
        cubeThreads{
            const double oneOrho = 1.0/r_Q[0]; 	   // inverse density
            const double u = r_Q[1]*oneOrho*vgeo;  // dimensional x-velocity
            const double v = r_Q[2]*oneOrho*vgeo;  // dimensional y-velocity
            const double w = r_Q[3]*oneOrho*vgeo;  // dimensional z-velocity
            const double p = (gm1)*(r_Q[4] - 0.5*r_Q[0]*(u*u + v*v + w*w);
            const double Ep = r_Q[4] + p;

            // convective flux: x-component
            s_fluxX[k][j][i][0] = r_Q[0]*(u);
            s_fluxX[k][j][i][1] = r_Q[1]*(u) + p*vgeo;
            s_fluxX[k][j][i][2] = r_Q[2]*(u);
            s_fluxX[k][j][i][3] = r_Q[3]*(u);
            s_fluxX[k][j][i][4] =     Ep*(u);
            // convective fluxes: y-component and z-component
        } @barrier("localMemFence");

        // residual per thread and update global residual memory: ops = 10*N^4
        cubeThreads {
            int idx1 = i*p_NF + j*tm_nf + k*tm_tm_nf + elem_idx;
            #pragma unroll p_NQP
            for (int qp = 0; qp < p_NQP; ++qp)
                #pragma unroll p_NF
                for(int n = 0; n < p_NF; ++n)
                    r_R[n] += s_dphiw[i][qp]*s_fluxX[k][j][qp][n];
                    r_R[n] += s_dphiw[j][qp]*s_fluxY[k][qp][i][n];
                    r_R[n] += s_dphiw[k][qp]*s_fluxZ[qp][j][i][n];

            #pragma unroll p_NF
            for(int n = 0; n < p_NF; ++n) R[idx1+n] = r_R[n];
        }
    }
}
\end{lstlisting} \label{OCCA-VOL}

\begin{figure}[ht]
\begin{center}\includegraphics[scale=0.25]{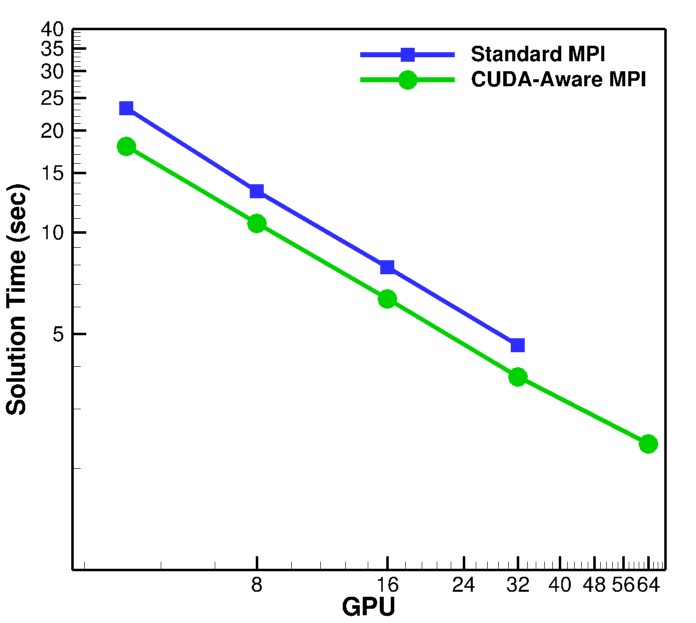}
\caption{Parallel strong scalability of CartDG on MIT Satori using up to 64 GPUs. A 29\% speedup is achieved at 4 GPUs contained in a single node and a 24\% speedup is achieved at 32 GPUs when using GPUDirect-enabled CUDA-Aware MPI compared to standard MPI.}
\label{fig:Satori}
\vspace{-1\baselineskip}
\end{center}
\end{figure}

\begin{table}[h]
\caption{CUDA-Aware MPI communication comparison to regular MPI communication using NVIDIA V100 GPUs.}
\begin{center}
  \begin{tabular}{c}
  \multicolumn{1}{c}{\textbf{MIT Satori: CUDA-Aware MPI}}\\
  \multicolumn{1}{c}{Problem Size: 452,984,832 DOF.}\\
  \begin{tabular}{|r|cl|l|}
      \hline
      \multicolumn{4}{|c|}{Solution Time (sec)} \\ \hline
         GPUs & ON    & OFF   & Speedup \\ \hline\hline
         4    & 17.94 & 23.22 & 1.29x   \\
         8    & 10.60 & 13.22 & 1.25x   \\
         16   & 6.34  & 7.87  & 1.24x   \\
         32   & 3.73  & 4.62  & 1.24x   \\
      \hline
 \end{tabular}
 \end{tabular}
\end{center}
\vspace{-1\baselineskip}
\label{Satori:CUDA-Aware}
\end{table}

\section{Performance Results: MIT Satori}
The GPU implementation of all computational kernels of CartDG is tested on the MIT Satori supercomputer. Satori is composed of 64 IBM Power9 nodes, each containing four NVIDIA V100 GPUs and two AC-922 22-core processors. The communication interconnect contains NVIDIA GPUDirect, which enables CUDA-Aware MPI communication protocols over 50 GB/s NVLink fabric linking the GPUs together within a node. The problem chosen for performance testing consists of a periodic domain containing $128\times128\times128$ mesh elements and solution polynomial degree of $p=6$, totaling 452,984,832 degrees-of-freedom (DOF) (2,264,924,160 unknowns).

Figure~\ref{fig:Satori} demonstrates the parallel strong scaling performance of the implementation comparing GPUDirect communication to non-CUDA-Aware MPI communication. As seen in the figure, the solver scales well to 16 nodes. Table~\ref{Satori:CUDA-Aware} demonstrates that the GPUDirect CUDA-Aware MPI provides a 29\% speedup using a single node and a 24\% speedup across multiple nodes.

\section{Performance Results: ORNL Summit}
We test the same implementation on ORNL Summit\footnote{These results were generated during early acceptence testing of ORNL Summit in 2018. GPUDirect MPI issues across multiple nodes were not fully resolved, thus, all multi-node results herein utilized non-CUDA-Aware MPI.}. Summit has nearly the same architecture as MIT Satori but contains 6 NVIDIA V100 GPUs and 2 IBM AC-922 22-core processors per node. Summit contains 4,608 nodes totaling 27,648 GPUs and 202,752 CPU cores and has achieved 148.6 PetaFLOPS in the LINPACK benchmark \cite{Dongerra:2003,Kahle:2019}. 

\subsection{Strong Scalability}
\begin{figure}[ht]
\begin{center}\includegraphics[scale=0.3]{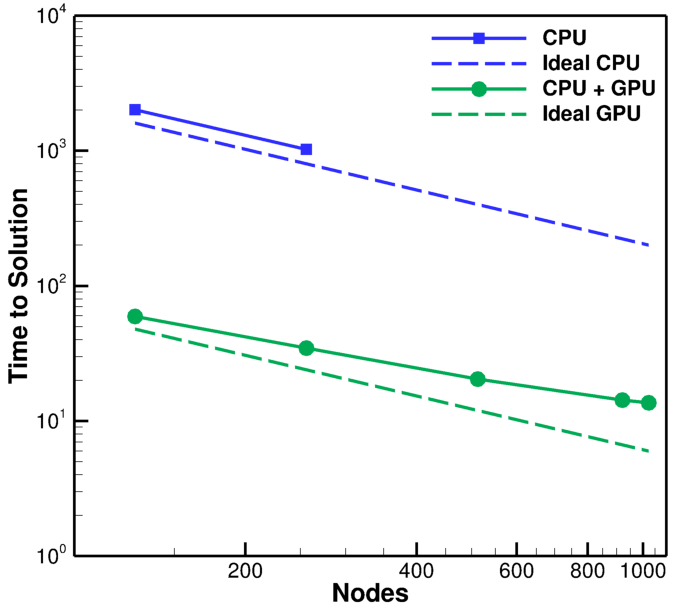}
\caption{Parallel strong scalability of CartDG on ORNL Summit using up to 1,024 nodes (6,144 GPUs/45,056 CPU cores).}
\label{Summit-Scaling}
\vspace{-1\baselineskip}
\end{center}
\end{figure}

We perform a second parallel strong-scalability test on ORNL Summit using a $512 \times 512 \times 768$ mesh composed of $p=6$, seventh-order accurate elements, totaling 69,055,021,056 degrees-of-freedom (345,275,105,280 unknowns). This problem size is chosen to give approximately 11 million degrees-of-freedom per GPU when using all GPUs on 1,024 nodes. The strong-scaling test measures the wall-clock time to solve 100 Low-Storage Explicit Runge-Kutta 5-stage 4th-order (LSERK45) time steps. The benchmark used 128, 256, 512, 920, and 1024 nodes. Figure~\ref{Summit-Scaling} displays the scaling results. We note that these results do not utilize CUDA-Aware MPI; the information required by neighboring cores is first transferred to the host CPU, exchanged via MPI, then transferred to the GPU. Table \ref{MPI-Summit} tabulates the wall-clock times at various node counts for the CPU and GPU implementations.

\begin{table}[h]
\caption{GPU kernel performance for 512 nodes (non-CUDA-Aware communication). Problem size: 69,055,021,056 degrees-of-freedom. (Theoretical peak assumed 21.5 PetaFlops for 512 Nodes).}
 \vspace{-2\baselineskip}
 \begin{center}
  \begin{tabular}{|r|rc|c|}
      \hline
      \multicolumn{4}{|c|}{\textbf{ORNL Summit: GPU Performance}} \\ \hline
         Kernel         & Time (sec) & PetaFlops & Achieved Theoretical Peak \\ \hline\hline
         Volume         & 1.77       & 4.84      & 22.5\%  \\
         Surface        & 2.15       & 2.45      & 11.4\%  \\
         Update-Project & 6.00       & 0.49      & 2.3\%   \\
      \hline
      Communication     & 10.18      & --        &  --     \\
      \hline\hline
      Overall           & 20.45      & 0.82      &  3.8\%  \\
      \hline
 \end{tabular}
  \vspace{-1\baselineskip}
\end{center}
\label{GPU-Breakdown}
\end{table}

\begin{table}[h]
 \vspace{-1\baselineskip}
\caption{CUDA-Aware MPI comparison to regular MPI on one node using six NVIDIA V100 GPUs demonstrating a 40\% speedup.}
\begin{center}
  \begin{tabular}{c}
  \multicolumn{1}{c}{\textbf{ORNL Summit: CUDA-Aware MPI}}\\
  \multicolumn{1}{c}{Problem Size: 21,337,344 DOF.}\\
  \begin{tabular}{|r|cl|}
      \hline
      \multicolumn{3}{|l|}{\ \ \ \ Single Node\ \ \ \ \ Time (sec)} \\ \hline
         Kernel         & ON    & OFF \\ \hline\hline
         Volume         & 1.67  & 1.68       \\
         Surface        & 2.04  & 2.04       \\
         Update-Project & 5.70  & 5.70       \\
      \hline
      Communication     & 0.65  & 5.13       \\       
      \hline\hline
      Overall           & 10.37 & 14.55      \\
      \hline
 \end{tabular}
 \end{tabular}
 \vspace{-2\baselineskip}
\end{center}
\label{CUDA-Aware}
\end{table}

\begin{table*}[h]
\centering
 \caption{ORNL Summit parallel strong scalability comparison of CPU vs GPU (non-CUDA-Aware communication).}
  \begin{tabular}{c}
  \multicolumn{1}{c}{\textbf{ORNL Summit: Parallel Strong Scalability}}\\
  \multicolumn{1}{c}{Problem Size: 69,055,021,056 degrees-of-freedom.}\\
  \multicolumn{1}{c}{Each node contains 2 IBM POWER9 processors (44 cores per node) and 6 NVIDIA V100 GPUs.}\\
  \begin{tabular}{|r|rc|}
      \hline
      \multicolumn{3}{|c|}{\textbf{CPU}}     \\ \hline
         Nodes  & Time (sec) & Scalability \% \\ \hline
         128    & 2011.2     & 100.0  \\
         256    & 1021.9     & 98.43 \\
         512    & --         & -- \\
         920    & --         & -- \\
       1,024    & --         & -- \\
      \hline
 \end{tabular}
 \begin{tabular}{|r|rc|}
      \hline
      \multicolumn{3}{|c|}{\textbf{GPU}}     \\ \hline
         Nodes  & Time (sec) & Scalability \% \\ \hline
         128    & 59.37  & 100.0 \\
         256    & 34.75  & 85.42 \\
         512    & 20.45  & 72.58 \\
         920    & 14.26  & 57.94 \\
       1,024    & 13.67  & 54.29 \\
      \hline
 \end{tabular}
 \begin{tabular}{|c|}
      \hline
      \multicolumn{1}{|c|}{\textbf{Speedup}}     \\ \hline
       CPU/GPU \\ \hline
       33.9$\mathbf{x}$ \\
       29.4$\mathbf{x}$ \\
       --      \\
       --      \\
       --      \\
      \hline
 \end{tabular}
 \end{tabular}
\label{MPI-Summit}
\end{table*}

\subsection{GPU Performance}
As demonstrated in the strong-scaling figure, the performance improvement is approximately 30x when using all six GPUs compared to using both CPUs per node only (using 44 cores per node). Table \ref{GPU-Breakdown} demonstrates the computational kernel breakdown of the GPU simulation on 512 nodes. The volume integration kernel corresponding to the code sample achieved nearly 5 PetaFLOPs of performance on 512 nodes. Additionally, the communication overhead is large in comparison to the computational time. This is due to not utilizing the CUDA-Aware MPI on Summit, which allows for faster GPU-GPU communication on a node and overlap of computation and communication. 

A communication benchmark is conducted using a single node with six GPUs interconnected with NVLink for a problem containing 21,337,344 DOFs. 
As shown in Table \ref{CUDA-Aware}, CartDG achieves an 8x speedup in communication time using the GPUDirect CUDA-Aware MPI on one node corresponding to a 40\% overall speedup in time to solution. This result corroborates the results achieved on MIT Satori and illustrates a significant point-to-point communication improvement when using GPUDirect CUDA-Aware MPI. This contrasts the result found by \cite{Romero:2020} which tested a similar high-order solver on ORNL Summit indicating a \textit{slow-down} in performance when using CUDA-Aware MPI.

\section{Conclusion}
This work developed of a high-order discontinuous Galerkin method on Cartesian meshes for the use of heterogeneous architectures including GPUs by porting all computational kernels via OCCA. Performance benchmarks on MIT Satori and ORNL Summit demonstrated significant speedups achieved including an overall 30x time to solution over the CPU-only implementation. Further, the benefits of utilizing GPUDirect for point-to-point communication were shown for this application, improving the overall performance by 24\% across multiple nodes and an 8x improvement in communication time within one node. If the CUDA-Aware MPI performance improvement holds at scale, as indicated by the results performed on MIT Satori, a GPU-to-CPU performance improvement of 40-50x may be expected.

\section*{Acknowledgment}
The authors acknowledge the MIT Research Computing Project and Oak Ridge National Laboratory for providing HPC resources that have contributed to the research results reported in this paper.


\begin{thebibliography}{00}
\bibitem{Hoefler:2019} T. Ben-Nun and T. Hoefler, ``Demystifying parallel and distributed deep learning: An in-depth concurrency analysis,'' ACM Computing Surveys (CSUR) 52, no. 4 (2019): 1-43.

\bibitem{Reuther:2019} A. Reuther, P. Michaleas, M. Jones, V. Gadepally, S. Samsi, and J. Kepner, ``Survey and benchmarking of machine learning accelerators," arXiv preprint arXiv:1908.11348 (2019).

\bibitem{Messina:2017} P. Messina, ``The exascale computing project,'' Computing in Science \& Engineering 19, no. 3 (2017): 63-67.

\bibitem{Jameson:1986} A. Jameson, and D. J. Mavriplis, ``Finite volume solution of the two-dimensional Euler equations on a regular triangular mesh,'' AIAA Journal 24, no. 4 (1986): 611-618.

\bibitem{Mavriplis:1988} D. J. Mavriplis, ``Multigrid solution of the two-dimensional Euler equations on unstructured triangular meshes,'' AIAA Journal 26, no. 7 (1988): 824-831.

\bibitem{Slotnick:2014} J. Slotnick, A. Khodadoust, J. Alonso, D. Darmofal, W. Gropp, E. Lurie, and D. J. Mavriplis, ``CFD vision 2030 study: a path to revolutionary computational aerosciences,'' NASA Technical Reports, 2014.

\bibitem{Hesthaven:2007} J. S. Hesthaven, and T. Warburton, \uline{Nodal discontinuous Galerkin methods: algorithms, analysis, and applications}. Springer Science \& Business Media, 2007.

\bibitem{Witherden:2014} F. D. Witherden, A. M. Farrington, and P. E. Vincent, ``PyFR: An open source framework for solving advection–diffusion type problems on streaming architectures using the flux reconstruction approach,'' Computer Physics Communications 185, no. 11 (2014): 3028-3040.

\bibitem{Klockner:2009} A. Klöckner, T. Warburton, J. Bridge, J. S. Hesthaven, ``Nodal discontinuous Galerkin methods on graphics processors,'' Journal of Computational Physics, 228(21), pp.7863-7882, 2009.

\bibitem{Gandham:2015} R. Gandham, D. Medina, and T. Warburton, ``GPU accelerated discontinuous Galerkin
methods for shallow water equations,'' Communications in Computational Physics, 18 (1), pp. 37–64, 2015.

\bibitem{Modave:2016} A. Modave, A. St-Cyr, and T. Warburton, ``GPU performance analysis of a nodal discontinuous Galerkin method for acoustic and elastic models,'' Computers \& Geosciences, 91, pp. 64–76, 2016.

\bibitem{Chan:2016} J. Chan, Z. Wang, A. Modave, J. F. Remacle, and T. Warburton, ``GPU-accelerated discontinuous Galerkin methods on hybrid meshes,'' Journal of Computational Physics, 318, pp. 142–168, 2016.

\bibitem{Karakus:2016a} A. Karakus, T. Warburton, M. H. Aksel, and C. Sert, ``A GPU-accelerated adaptive
discontinuous Galerkin method for level set equation,'' International Journal of Computational Fluid Dynamics, 30 (1), pp. 56–68, 2016.

\bibitem{Karakus:2016b} A. Karakus, T. Warburton, M. H. Aksel, and C. Sert, ``A GPU accelerated level set reinitialization for an adaptive discontinuous Galerkin method,'' Computers \& Mathematics with Applications, 72 (3), pp. 755–767, 2016.

\bibitem{Karakus:2019} A. Karakus, N. Chalmers, K. Świrydowicz, and T. Warburton, ``A GPU accelerated discontinuous Galerkin incompressible flow solver,'' Journal of Computational Physics, 390, pp.380-404, 2019.

\bibitem{Medina:2014} D. S. Medina, A St-Cyr, and T. Warburton, ``OCCA: A unified approach to multi-threading languages,'' arXiv preprint arXiv:1403.0968 (2014).

\bibitem{Medina:2015} D. Medina, ``OKL: A unified language for parallel architectures,`` Ph. D. Dissertation, Rice University, 2015.

\bibitem{Abdi:2019a} D. S. Abdi, L. C. Wilcox, T. C. Warburton, and F. X. Giraldo, ``A GPU-accelerated continuous and discontinuous Galerkin non-hydrostatic atmospheric model,'' The International Journal of High Performance Computing Applications, 33(1), pp.81-109, 2019.

\bibitem{Abdi:2019b} D. S. Abdi, F. X. Giraldo, E. M. Constantinescu, L. E. Carr, L. C. Wilcox, and T. C. Warburton, ``Acceleration of the IMplicit–EXplicit nonhydrostatic unified model of the atmosphere on manycore processors,'' The International Journal of High Performance Computing Applications, 33(2), pp.242-267, 2019.

\bibitem{Cockburn:2012} B. Cockburn, G. E. Karniadakis, and C. Shu, \uline{Discontinuous Galerkin methods: theory, computation and applications}. Vol. 11. Springer Science \& Business Media, 2012.

\bibitem{Harten:1983} A. Harten, P. D. Lax, and B. van Leer, ``On upstream differencing and Godunov-type schemes for hyperbolic conservation laws,'' SIAM review 25, no. 1 (1983): 35-61.

\bibitem{Brazell:2016} M. J. Brazell, A. C. Kirby, J. Sitaraman, and D. J. Mavriplis, ``A multi-solver overset mesh Approach for 3D mixed element variable order discretizations,'' AIAA Paper 2016-0053, 54th AIAA Aerospace Sciences Meeting, San Diego CA, January 2016.

\bibitem{Kirby:2017} A. C. Kirby, and D. J. Mavriplis, ``High Fidelity Blade-Resolved Wind Plant Modeling.'' SC17, 2017.

\bibitem{Kirby:2018a} A. C. Kirby, Z. Yang, D. J. Mavriplis, E. P. Duque, and B. J. Whitlock, ``Visualization and data analytics challenges of large-scale high-fidelity numerical simulations of wind energy applications,'' AIAA Paper 2018-1171, 56th AIAA Aerospace Sciences Meeting, Kissimmee FL, January 2018.

\bibitem{Kara:2020} K. Kara, A. C. Kirby and D. J. Mavriplis, ``Hover Predictions of S76 Rotor Using a High Order Discontinuous Galerkin Off-Body Discretization,'' AIAA Paper 2020-0771 Presented at the AIAA Scitech 2020 Conference, Orlando FL, January 2020.

\bibitem{Kirby:2018} A. C. Kirby, A. Hassanzadeh, D. J. Mavriplis, and J. W. Naughton, ``Wind Turbine Wake Dynamics Analysis Using a High-Fidelity Simulation Framework with Blade-Resolved Turbine Models,'' AIAA Paper 2018-0256, Wind Energy Symposium, Kissimmee FL, January 2018.

\bibitem{Kirby:2019} A. C. Kirby, M. J. Brazell, Z. Yang, R. Roy, B. R. Ahrabi, M. K. Stoellinger, J. Sitaraman, and D. J. Mavriplis, ``Wind farm simulations using an overset hp-adaptive approach with blade-resolved turbine models,'' The International Journal of High Performance Computing Applications 33, no. 5 (2019): 897-923.

\bibitem{Kirby:2016} A. Kirby, M. Brazell, J. Sitaraman, D. Mavriplis, ``An Overset Adaptive High-Order Approach for Blade-Resolved Wind Energy Applications,'' AHS Forum 72, West Palm Beach FL, May 2016.

\bibitem{Edmonds:2019} A. P. Edmonds, A. Hassanzadeh, A. C. Kirby, D. J. Mavriplis, and J. W. Naughton, ``Effects of Blade Load Distributions on Wind Turbine Wake Evolution Using Blade-Resolved Computational Fluid Dynamics Simulations,'' AIAA Paper 2019-2081, 57th AIAA Aerospace Sciences Meeting, San Diego CA, January 2019.

\bibitem{Mavriplis:2015} D. Mavriplis, M. Long, T. Lake and M. Langlois, ``NSU3D results for the second AIAA high-lift prediction workshop,'' Journal of Aircraft 52, no. 4 (2015): 1063-1081.

\bibitem{Stoellinger:2019} M. K. Stoellinger, A. P. Edmonds, A. C. Kirby, D. J. Mavriplis and S. Heinz, ``Dynamic SGS modeling in LES using DG with kinetic energy preserving flux schemes,'' AIAA Paper 2019-1648, 57th AIAA Aerospace Sciences Meeting, San Diego CA, January 2019.

\bibitem{Brazell:2016:tioga} M. J. Brazell, J. Sitaraman, and D. J. Mavriplis, ``An overset mesh approach for 3D mixed element high-order discretizations,'' Journal of Computational Physics 322 (2016): 33-51.

\bibitem{Sitaraman:2020} J. Sitaraman, D. Jude, and M. J. Brazell, ``Enhancements to Overset Methods for Improved Accuracy and Solution Convergence,'' AIAA Paper 2020-1528, AIAA Scitech Forum, 2020.

\bibitem{Kirby:2015} A. C. Kirby, D. J. Mavriplis, and A. M. Wissink, ``An Adaptive Explicit 3D Discontinuous Galerkin Solver For Unsteady Problems,'' AIAA Paper 2015-3046, 22nd AIAA Computational Fluid Dynamics Conference, Dallas TX, June 2015.

\bibitem{Brazell:2015} Ma. J. Brazell, M. J. Brazell, M. K. Stoellinger, D. J. Mavriplis, and A. C. Kirby, ``Using LES in a Discontinuous Galerkin method with constant and dynamic SGS models,'' AIAA Paper 2015-0060, 53rd AIAA Aerospace Sciences Meeting, Kissimmee FL, January 2015.

\bibitem{Kirby:thesis}
A. C. Kirby, ``Enabling High-order Methods for Extreme-scale Simulations,'' University of Wyoming, 2018.

\bibitem{Brazell:2017} M. J. Brazell, A. C. Kirby, and D. J. Mavriplis, ``A High-order Discontinuous-Galerkin Octree-Based AMR Solver for Overset Simulations,'' AIAA Paper 2017-3944, 23rd AIAA Computational Fluid Dynamics Conference, Denver CO, June 2017.

\bibitem{Dongerra:2003} J. J. Dongarra, P. Luszczek, and A. Petitet, ``The LINPACK benchmark: past, present and future,'' Concurrency and Computation: practice and experience 15, no. 9 (2003): 803-820.

\bibitem{Kahle:2019} J. A. Kahle, J. Moreno, and D. Dreps, ``2.1 Summit and Sierra: Designing AI/HPC Supercomputers,'' In 2019 IEEE International Solid-State Circuits Conference-(ISSCC), pp. 42-43. IEEE, 2019.

\bibitem{Romero:2020} J. Romero, J. Crabill, J. E. Watkins, F. D. Witherden, and A. Jameson, ``ZEFR: A GPU-accelerated high-order solver for compressible viscous flows using the flux reconstruction method,'' Computer Physics Communications (2020): 107169.

\end{thebibliography}
\end{document}